\documentclass{article}
\usepackage{newpasp,graphicx}
\def\edcomment#1{\iffalse\marginpar{\raggedright\sl#1\/}\else\relax\fi}
\reversemarginpar

\begin{document}
\title{A multiwavelength vision of star forming regions: What does $\lambda_1$ know about $\lambda_2$?}
 \author{Miguel Cervi\~no}
\affil{Instituto de Astrof\'\i sica de Andaluc\'\i a (CSIC), Apdo. 3004, 
 18008 Granada, Spain}
\affil{Laboratorio de Astrof\'\i sica Espacial y F\'\i sica Fundamental
 (INTA), Apdo. 50727, 28080 Madrid, Spain}

\begin{abstract}
In this contribution I show the physical limitations to the application of
standard synthesis models from 0 to 10 Gyr. I also present the
multiwavelength spectrum (from $\gamma$-rays to radio) of young star
forming regions based on evolutionary population synthesis models that
takes into account the stochasticity of the star formation proces. I show
how the correlation of different observables can allow to establish a
better understanding of the star formation processes. Finally I compare the
spectral energy distribution of AGNs and star forming regions and I show
the possible connection of Star Forming Galaxies with Seyfert 2 AGNs.

\end{abstract}

\section{Introduction: Intrinsic limits of synthesis models}

Since the work of Tinsley \& Gunn (1976), evolutionary synthesis models have
been extensively used to obtain the integrated properties of systems where
stars are formed.  Models are based on the convolution of isochrones with the
Initial Mass Function (IMF) and the Star Formation History. For the case of
an instantaneous burst of star formation (simple stellar population, SSP),
the {\it mean luminosity} in a given band and a given age, $\mu_L(t)$,
results from the weighted sum of the number of stars with initial mass
$m_i$, $w_i$, (given by the IMF), and the individual luminosities
$l_i(m_i,t)$ (given by the isochrone).  If the sum of the $w_i$ values is
normalized to 1 M$_\odot$ transformed into stars from the onset of the burst
(as usual), the resulting luminosity will also be normalized.  The total
luminosity of a cluster is then, directly proportional to the initial mass
transformed into stars, $\cal M$: $L^{clus}(t) = {\cal M} \times \mu_L(t)$.

Such a modeling has some intrinsic constraints: 
\begin{itemize}
\item The total luminosity of the cluster modeled, $L^{clus}(t)$, must be
larger than the individual contribution of any of the stars included in the
model, and, in particular, larger than the most luminous star,
$l_i^{max}(m_i,t)$.  This statement defines a natural theoretical limit,
that is not always considered when the models are applied to real
observations. The modeling of clusters with masses below this limit can
only be performed by codes that include sampling effects, either
analytically or via Monte Carlo simulations (see, e.g. Cervi\~no et
al. 2002).

\item There is also an uncertainty coming from the very nature of the
IMF. Most theoretical models assume that the IMF is completely populated,
but Nature does not follow such rule. In other words, any modeling that
assumes a completely populated IMF will be correct only under the
asymptotic assumption of an infinite number of stars. Otherwise, the
modeling only yields a mean value of the observed quantities. In fact,
Monte Carlo simulations for clusters at different ages show that the mean
values of the synthesized quantities depend on the amount of stars used for
the simulations (see Santos \& Frogel 1997, Cervi\~no, Luridiana \&
Castander 2000 or Bruzual 2002 as examples). The main difference between
Monte Carlo and standard simulations is that Monte Carlo simulations always
use an integer number of stars.  On the other hand, by construction,
standard (analytical) simulations always assume that most of the mass
values are represented by a fractional number of stars.

Note that such uncertainty is relevant to any modeling that makes use
of the IMF. An example of its influence in galactic chemical evolution
models can be found in Cervi\~no \& Moll\'a (2002).

\end{itemize}

\begin{figure}[tbh]
\centering \includegraphics[width=\textwidth]{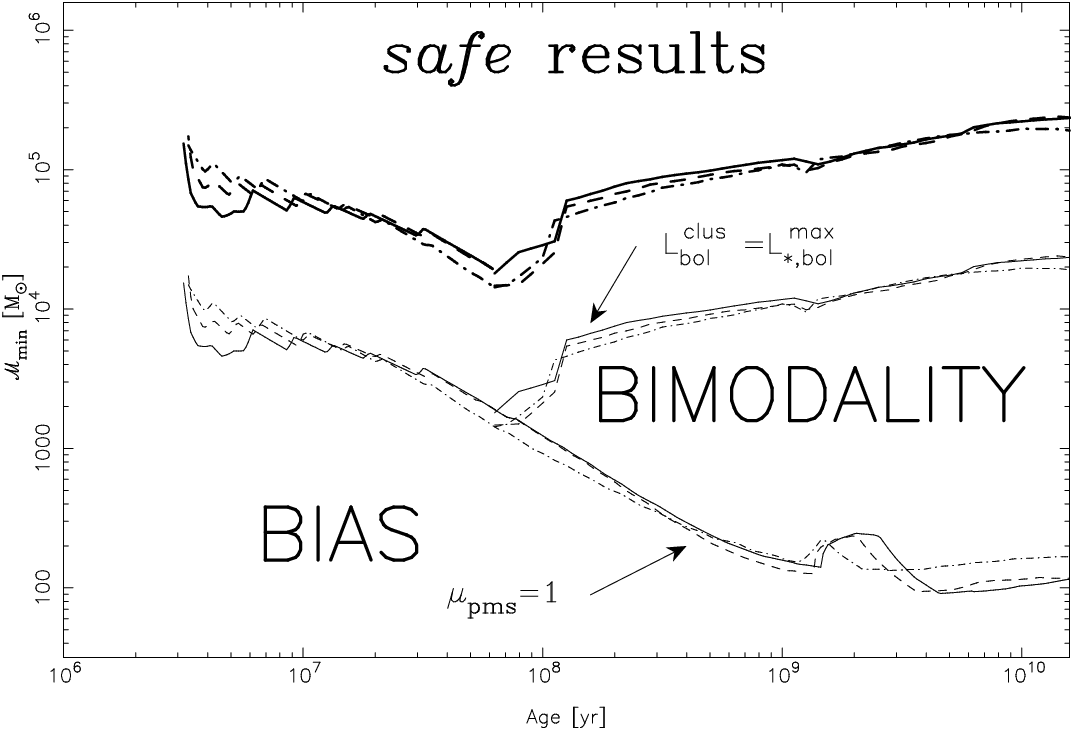}
\caption{${\cal M}$ ranges where different situations may occur when the
results of synthesis models are compared with observed data as a function
of time and metallicity: Z=0.019 (solid line), Z=0.008 (dashed line) and
Z=0.0004 (point-dashed line). The upper line defines the ${\cal M}$ value
where there are, at least, 10 post-main sequence stars ($t < 10^8$ yr) or
where $L_{bol}^{clus} > 10 \times l_{bol}^{max}(m_i,t)$ ($t > 10^8$).}
\label{fig:Mlim}
\end{figure}

Then, before using a synthesis model, it is necessary to know the ${\cal
M}$ value for which a variation of $\pm 1$ star is relevant to the results.
This value can be estimated imposing that $L_{bol}^{clus}$ is larger than
$10 \times l_{bol}^{max}(m_i,t)$. The resulting values for $\cal M$ for a
Salpeter IMF in the mass range 0.08 -- 120 M$_\odot$ are shown in Fig. 1
using the results from Girardi et al. (2000)\footnote{Available at {\tt
http://pleiadi.pd.astro.it/}}. It yields a value of ${\cal M} > 10^5$
M$_\odot$, that assure that there are, at least, 10 Post-main sequence
stars ($t < 10^8$ yr) or that $L_{bol}^{clus} > 10 \times
l_{bol}^{max}(m_i,t)$ ($t > 10^8$).

\section{Statistical synthesis models}

The only way to deal with small systems is to include statistical effects
in the synthesis models. This can be performed computing the variance of
the observable, $\sigma^2_L(t)$, together with the mean value. It can be
done easily assuming that the $w_i$ values follow a Poissonian distribution
(they must always be positive and integer numbers, see Cervi\~no et
al. 2002 for details). However, it is more useful to show the variance in
terms of an {\it effective number} of stars, ${\cal N}_L(t) =
{\mu^2_L(t)}/{\sigma^2_L(t)}$ which was defined by Buzzoni (1989). Both,
variance and $\cal N$, scale linearly with $\cal M$: the smaller the
system, the smaller $\cal N$ (and the larger the relative error). This
formulation has the additional advantage of enabling us to obtain the
correlation coefficient between two observables, $\rho(L_1,L_2)$.  This
coefficient is independent of $\cal M$ and it only depends on the IMF
slope, so it can also be used for the analysis of large samples (see the
contribution from R. Terlevich on the subject of mega datasets).  Finally,
the use of $\cal N$ also enables us to know when multi- or bi-modal
distributions may be observed and when the results of synthesis models are
biased in comparison with the observations (see Cervi\~no \& Valls-Gabaud
2002 for details).

In the following I am going to apply our synthesis models results to
different astrophysical scenarii. Model outputs\footnote{Not all the model
output are presented in the {\it www} server, please, ask us for any special request.} can be found
in the {\it www} server
\begin{center}
{\tt http://www.laeff.esa.es/users/mcs/SED}
\end{center}

\subsection{Multiwavelength emission}

In Fig.2 I show the multiwavelength spectrum of a 5.5 Myr old burst of star
formation including the 90\% confidence interval that arise from sampling
fluctuations in the stellar population for clusters with a given $\cal M$
and a Salpeter IMF slope with mass limits 2 -- 120 M$_\odot$. It is
interesting to note that the lower dispersion corresponds to the UV
continuum, which turns out to be the most reliable age indicator (in
absence of extinction effects).

Additionally, the correlation coefficients obtained theoretically can allow
to establish which are the better observables to determine physical
properties. Let me give you an example: The presence of Wolf-Rayet (WR)
stars imply the presence of ionizing flux, hence, some degree of
correlation is expected in the age-diagnostic diagrams of EW(H$\beta$)
vs. EW(WR). The age-dependent correlation coefficient defines the geometry
and the way how the observational data must be interpreted in such a
plot, in the sense that variations in one of the observables also determine
how the other observable varies. Such variations are only dependent on the
age and the IMF slope since the correlation coefficient is sampling
independent. 

\begin{figure}[tbh]
\centering \includegraphics[width=\textwidth]{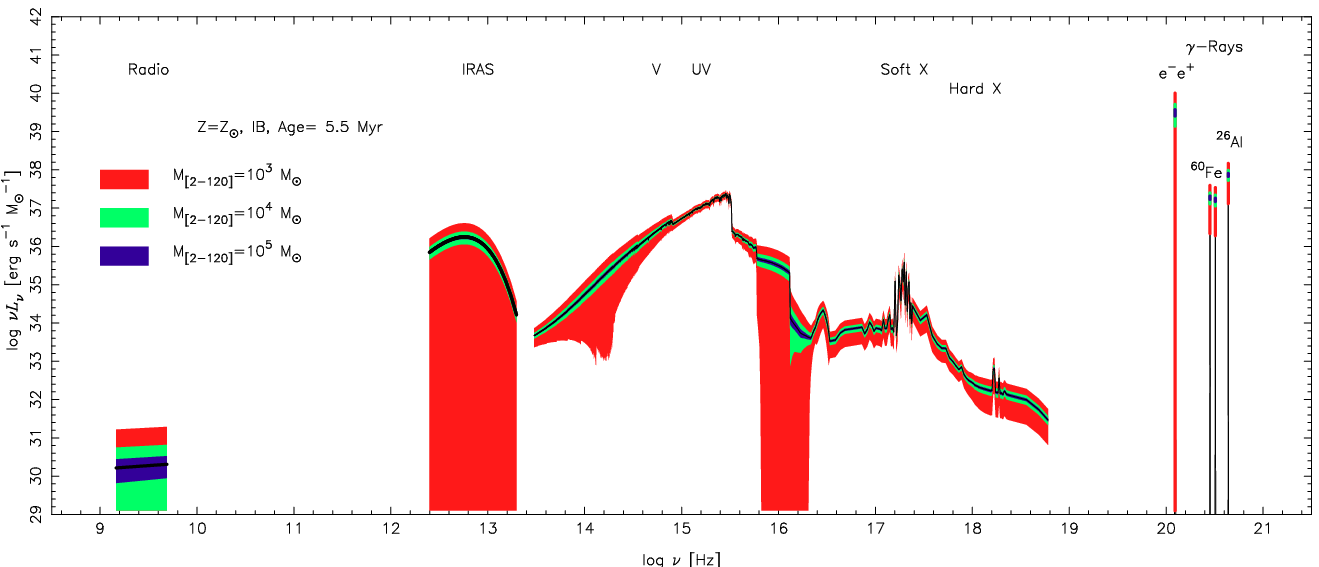}
\caption{Analytical 90\% Confidence Level for the multiwavelength spectrum
for a 5.5 Myr old burst.}
\end{figure}

Finally, there remains the problem of how to apply synthesis models to
individual small systems. In this case, the situation is more
complicated. Taking again the example of WR stars, whereas the presence of
such stars enable us to quote an age range where the star formation began,
their absence cannot be used to quote any plausible age range.  However, if
synthesis models have a physical reality, different observables must be
related to each other. Hence, the real solution will lie in the joint
distribution of the probability distributions of the different
observables. Unfortunately, this kind of study has not yet been performed.

\subsection{Starburts-AGN connection}

In the preceding items I have shown that synthesis models must be applied
with cautions to small systems. However, the definition of {\it small}
depends on the statistics of the observable. In Cervi\~no, Mas-Hesse \&
Kunth (2002) we have computed the soft X-ray emission from starburst
galaxies. This emission is produced from the reprocessing of the kinetic
energy produced by the star forming regions and from the Supernova
Remnants.  In this case, the dispersion arises mainly from the occurrence
of Supernova (SN) events, that is intrinsically a poor statistical
observable: the SN rate is around 10$^{-9}$ SN yr$^{-1}$ M$_\odot^{-1}$
(following a Salpeter IMF in the mass range 2 -- 120 M$_\odot$, Cervi\~no
\& Mas-Hesse 1994), and the X-ray life time of an event is around 10$^4$
yr, which implies $\cal M$ larger than 10$^5$ in order to have a {\it mean}
value of 1 SN active in X-rays. In that work the X-ray emission of
individual SN events was averaged over the SN life time (see Aretxaga et
al. in this proceedings for the statistics and evolution of SN fluxes),
hence the quoted values are an average estimation with an additional
dispersion that has not been considered. This additional dispersion may be
relevant in the interpretation of the optical emission line spectrum in
star forming galaxies (see also Rodr\'\i guez-Gaspar \& Tenorio-Tagle 1998,
Silich et al 2001, and Stasi\'nska \& Izotov 2002)

Besides the limitations, the approach is enough to test the relevance of
circumnuclear star forming regions in the energy budget of Active Galactic
nuclei (AGN). In Fig. 3, left panel, I show the soft-X ray/UV ratio
predicted by the synthesis models and their comparison with the
observational data of Star forming galaxies. In all the cases, some amount
of kinetic energy must be reprocessed in X-ray in order to explain the
observations. The comparison with Seyfert galaxies is shown in the right
panel of the figure. Star forming regions are able to explain the X-ray
emission of some Seyfert 2 galaxies, but their emission is not enough to
explain the X-ray emission in Seyfert 1. In this approach, the putative
black hole in some Seyfert 2 galaxies will only be relevant in the hard
X-ray domain.

\begin{figure}[tbh]
\centering \includegraphics[width=\textwidth]{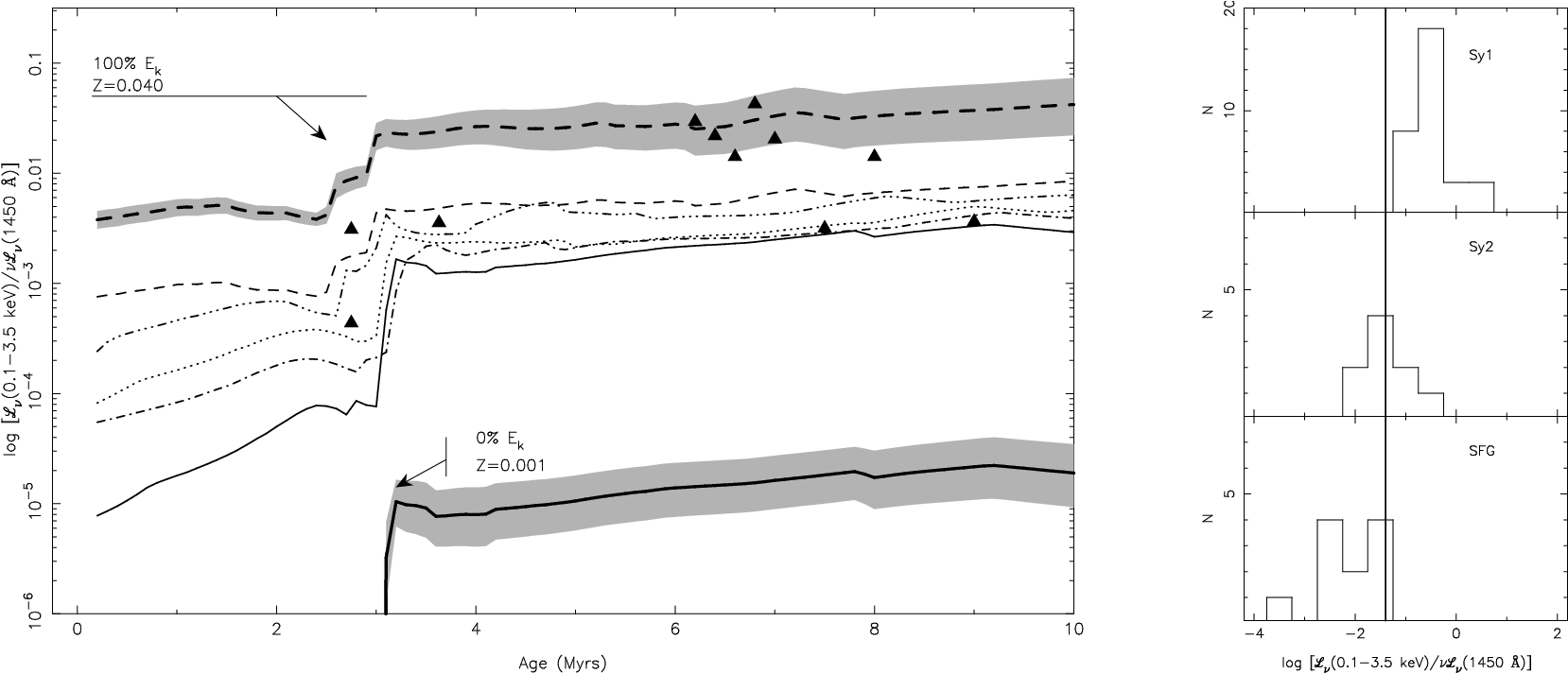}
\caption{{\it Left:} Ratio L$_\nu$(0.1 -- 3.5 keV) over $\nu$L$_\nu$(1450
\AA ) computed for a 20\% reprocessing values of mechanical energy into
soft X rays and different metallicities. In grey, the 90\% CL for a cluster
where $10^5$ M$_\odot$ have been transformed into stars for two extreme
metallicities and reprocessing values.  Data points from star forming
galaxies are marked as triangles.  {\it Right:} Histograms for the ratio
X-ray/UV for different types of emission line galaxies. The vertical solid
line corresponds to the upper value predicted by our evolutionary synthesis
models.}
\end{figure}

\acknowledgments I am gratefully for useful discussions with V. Luridiana,
J.M. Mas-Hesse, E. P\'erez, D. Valls-Gabaud, J.M. V\'\i lchez,
G. Stasi\'nska and D. Kunth at different stages of this project.  This
project has been partially supported by the AYA 3939-C03-01 program. I also
thank the LOC for financial support.

%\clearpage

\section*{\bf Discussion}

\noindent
{\it Dottori:} How circumnuclear are the circumnuclear star forming regions you
are talking about?\\

\noindent
{\it Cervi\~no:} The data I have presented are based on {\it IUE}
observations, so they are in a region of $10\times20$ arc sec around the
center.\\

\noindent
{\it Cid Fernandes:} You have rightfully warned us that evolutionary synthesis
predictions come with a statistical uncertainty. Have you compared your
predictions for the variances with the observed dispersion in properties of
star-forming regions?\\

\noindent
{\it Cervi\~no:} Not personally. At this moment I am still working on the
statistical implications. However, other authors (Bresolin \& Kennicutt
2002) have applied the models results available on the Web server to
high-metallicity star-forming regions and the theoretical dispersion looks
to be consistent with the observed data.\\

\noindent
{\it Garc\'\i a Vargas:} It is true that taking into account all the
probabilistic effects on cluster modeling is important to understand the
dispersion, or degree of correlation when plotting observed data of
GEHRs. However I think that this approach should be systematically applied
to large GEHR where there is evidence of several small clusters ($< 10^5$
M$_\odot$) from the region (as revealed from {\it HST})\\

\noindent
{\it Cervi\~no:} Yes, indeed the effect of sampling is more dramatic for low
mass clusters (${\cal M} < 10^5$ M$_\odot$), but it will be also present
in more massive systems.\\

\end{document}